\documentclass[twocolumn]{aastex63}
\usepackage[perpage,symbol*]{footmisc}
\usepackage{color}
\usepackage{graphicx}
\usepackage{amsmath,bm}
\newcommand{\be}{\begin{equation}}
\newcommand{\ee}{\end{equation}}
\newcommand{\bq}{\begin{eqnarray}}
\newcommand{\eq}{\end{eqnarray}}

\newcommand{\n}{\nonumber}
\def\({\left(}
\def\){\right)}
\def\[{\left[}
\def\]{\right]}

\submitjournal{ApJ}
\shorttitle{Model-independent constraints on $\Omega_k$}
\shortauthors{Bo Wang et al.}

\begin{document}
\title{Cosmological model-independent constraints on spatial curvature from strong gravitational lensing and type Ia supernova observations}

\correspondingauthor{Jing-Zhao Qi, Xin Zhang}
\email{qijingzhao@mail.neu.edu.cn, zhangxin@mail.neu.edu.cn}

\author{Bo Wang}
\affiliation{Department of Physics, College of Sciences, Northeastern
University, Shenyang 110819, China}

\author{Jing-Zhao Qi}
\affiliation{Department of Physics, College of Sciences, Northeastern
University, Shenyang 110819, China}

\author{Jing-Fei Zhang}
\affiliation{Department of Physics, College of Sciences, Northeastern
University, Shenyang 110819, China}

\author{Xin Zhang}
\affiliation{Department of Physics, College of Sciences, Northeastern
University, Shenyang 110819, China}
\affiliation{Ministry of Education's Key Laboratory of Data Analytics and Optimization
for Smart Industry, Northeastern University, Shenyang 110819, China}
\affiliation{Center for High Energy Physics, Peking University, Beijing 100080, China}

\begin{abstract}
Applying the distance sum rule in strong gravitational lensing (SGL) and type Ia supernova (SN Ia) observations, one can provide an interesting cosmological model-independent method to determine the cosmic curvature parameter $\Omega_k$. In this paper, with the newly compiled data sets including 161 galactic-scale SGL systems and 1048 SN Ia data, we place constraints on $\Omega_k$ within the framework of three types of lens models extensively used in SGL studies. Moreover, to investigate the effect of different mass lens samples on the results, we divide the SGL sample into three sub-samples based on the center velocity dispersion of intervening galaxies. In the singular isothermal sphere (SIS) and extended power-law lens models, a flat universe is supported with the uncertainty about 0.2, while a closed universe is preferred in the power-law lens model.  We find that the choice of lens models and the classification of SGL data actually can influence the constraints on $\Omega_k$ significantly.
\end{abstract}


\section{Introduction} \label{introduction}
As an important cosmological parameter, the spatial curvature parameter $\Omega_k$ of the universe determines whether the space of the universe is open, flat, or closed, which is intimately related to the evolution of the universe. From the precise measurements of the anisotropies of the cosmic microwave background (CMB) by the Planck satellite mission \citep{ade2016planck}, the constraint result of $|\Omega_k|<0.005$ is derived, showing that the space of the universe is preferred to be flat.

However, for this tight constraint result of $\Omega_k$, two points should be noted: (1) This is a cosmological model-dependent measurement result. A specific cosmological model, i.e., the cosmological constant plus cold dark matter model, usually abbreviated as $\Lambda$CDM model, which is often viewed as the standard model of cosmology, is assumed for the derivation of this result. (2) This result is based on an early-universe measurement, because the CMB originates from the photons' decoupling in the early universe.

Although the Planck mission's measurements combined with the $\Lambda$CDM model can precisely determine some cosmological parameters (in particular for the six base parameters), some problems still have been puzzling the current cosmology. For example, among these problems the most prominent one is the ``Hubble tension'' problem, i.e., the measurement results of the Hubble constant $H_0$ from the Planck observation and the distance-ladder observation are rather inconsistent, with the tension between them being 4.4$\sigma$ significance \citep{riess2019large}. In addition, although the accelerated expansion of the universe is confirmed by the current observations, the true physics behind it is still an open question: We still cannot confirm that dark energy is a cosmological constant, and actually a dynamical dark energy or a modification of gravity on cosmological scales can also be responsible for the cosmic acceleration.

{In addition, the recent Planck 2018 results have confirmed the presence of a higher lensing amplitude, $A_{\mathrm{lens}}$, in CMB power spectra compared to the one predicted in the $\Lambda$CDM model at about 3$\sigma$ level \citep{Aghanim:2018eyx,Aghanim:2019ame}. By extending the base $\Lambda$CDM model to include the spatial curvature parameter, \citet{DiValentino:2019qzk} obtained the result of a closed universe at 3.4$\sigma$ level, $-0.095<\Omega_k <-0.007$ (99\% CL), using the temperature and polarization angular power spectra of Planck 2018. Since the curvature parameter $\Omega_k$ is positively correlated with $A_{\mathrm{lens}}$, the result consistent with $A_{\mathrm{lens}}=1$ can be obtained naturally in a closed universe, thus providing a solution to the $A_{\mathrm{lens}}$ anomaly of CMB lensing. This further exacerbates the inconsistencies between cosmological observations. Thus, the tension between early-universe and late-universe measurements and the mystery of dark energy urge us to re-examine the result of the cosmic curvature through a cosmological model-independent method and using only low-redshift observations.}

As early as about a decade ago, \citet{Clarkson:2007pz} proposed a cosmological model-independent method to measure the cosmic curvature using standard rulers or standard candles, and to directly test the Friedmann-Lemaiter-Robertson-Walker (FLRW) metric, which has been fully implemented with current observational data including type Ia supernovae (SN Ia, or simply SN) (standard candles) and the Hubble parameter $H(z)$ (cosmic chronometers) \citep{Li:2016wjm,Wei:2016xti}. The cosmic curvature estimated in this way is well consistent with a flat universe with $\Omega_k=0$.

On the other hand, based on the distance sum rule, \citet{rasanen2015new} realized that a joint analysis with the strong gravitational lensing (SGL) and SN Ia data may provide a test for the validity of the FLRW metric by comparing whether the model-independent measurements of $\Omega_k$ at different redshifts are equal. However, \citet{qi2019strongly} argued that this method makes a strong assumption based on the isotropy and homogeneity of the universe in fact, and accordingly they proposed a valid test for the FLRW metric through the multiple measurements of SGL systems with SN Ia as background sources. Actually, the method proposed by \citet{rasanen2015new} can be used to measure the cosmic curvature parameter without assuming any cosmological models, and in their paper they found that the cosmic curvature is close to zero with a prior from the CMB observation and the local measurement of the Hubble constant. Following that, \citet{xia2017revisiting} obtained similar results using an updated SGL sample including 118 galactic-scale systems \citep{cao2015cosmology} and considering more complex lensing models in an attempt to reduce the systematic uncertainty. Subsequently, \citet{li2018curvature} updated the constraint on $\Omega_k$ with the joint light-curve analysis (JLA) SN Ia sample \citep{betoule2014improved}, and found that a spatially closed universe is preferred. They also speculate that there are some unknown systematics leading to a bias in estimating the cosmic curvature parameter from SGL observations, and expect a larger SGL sample from the near future survey telescope to clarify this issue.

More recently, by replacing the SN Ia data with the VLBI observations of milliarcsecond compact structure in intermediate-luminosity quasars at higher redshifts up to 2.76, \citet{qi2018distance} found that the constraint on the cosmic curvature is marginally compatible with the flat universe case. Besides, they also found that the constraints on $\Omega_k$ from SGL observations are different in three sub-samples defined according to the lens velocity dispersion, which implies that it is actually not reasonable to characterize all lenses with a uniform model. Therefore, actually, to accurately measure the spatial curvature of the universe through SGL observations within the framework of distance sum rule is still on the way.

Fortunately, the latest larger and higher redshift observations provide an opportunity for us to explore this issue further. First, the Pan-STARRS1 Medium Deep Survey released a new sample of SN Ia data including 1048 SN in the redshift range $0.001<z<2.3$, named the Pantheon sample \citep{scolnic2018complete}. Compared with the JLA sample consisting of 740 SN in the redshift range $0<z<1.3$, one can find that both the sample size and the high redshift data have been  improved significantly in the Pantheon sample, which means that the number of the SGL systems that we can calibrate the distance by using SN data is increased. Second, \citet{Chen:2018jcf} recently compiled the largest SGL sample so far including 161 galaxy-scale strong lensing systems with both high resolution imaging and stellar dynamical data, in which the redshift range of lens galaxies is $0.0625<z_l<0.958$ and the sources are extended to $0.196<z_s<3.595$.

On the other hand, in most of previous studies, the methods using the SN Ia data to calibrate the distances of lenses and sources in SGL systems are usually based on fitting a simple third (or fourth)-order polynomial function. However, it is difficult to clarify whether the polynomial function used can accurately describe the relation between the cosmological distance and the redshift, and whether it would bring extra bias. Therefore, in this work, we use the method of Gaussian Process to reconstruct a smooth distance-redshift curve directly from the SN Ia data without assuming any parametric forms or cosmological models, and then we calibrate the distances of lenses and sources in SGL systems.

In this paper, with these three improvements, we will update the constraints on the cosmic curvature parameter based on the distance sum rule. In addition, according to the clues from \cite{qi2018distance}, we will also divide the sample into three sub-samples, i.e., the low-, intermediate-, and high-mass sub-samples based on lens velocity dispersion, and make an analysis for them to study the effect of lens samples on the measurement of the cosmic curvature.

This paper is organized as follows. In Section \ref{sec2}, we introduce the methodology used in this work including the distance sum rule and Gaussian Process. In this section, we also introduce the observational data, including the Pantheon SN Ia sample and the newly compiled SGL sample with three lens models extensively used in SGL studies. The results obtained using different lens models and subsamples are presented in Section \ref{sec3}. Finally, the conclusions are given in Section \ref{sec4}.

\section{Methods and data}
\label{sec2}

\subsection{Distance sum rule}
According to the cosmological principle, i.e., the universe is homogeneous and isotropic on large scales, the geometry of the universe is described by the FLRW metric as
\begin{equation}
d s^2=-d t^2+\frac{a^2(t)}{1-k r^2}d r^2+{a^2(t)} r^2 d \Omega^2, \label{1}
\end{equation}
where $k$ is a constant representing the spatial curvature and is related to the curvature parameter as $\Omega_{k}=-k/{H^{2}_{0}}$, with $H_0$ being the Hubble constant. Here, $\Omega_k<0$, $\Omega_k>0$, and $\Omega_k=0$ correspond to the closed, open, and flat universe cases, respectively.

Considering a strong lensing system with the intervening galaxy at redshift $z_l$ acting as a lens, the separation of multiple images of source at redshift $z_s$ depends on the angular diameter distance ratio $D_A(z_l,z_s)/D_A(z_s)$ as long as that one provides a reliable lens model for the mass distribution. The dimensionless comoving distance $d(z)$ between the lens and the source can be expressed as
\begin{eqnarray}
d(z_l,z_s)&=& (1+z_s)H_0D_A(z_l,z_s)  \nonumber \\
&=& \frac{1}{\sqrt{|\Omega_k|}}{\rm sinn}[\sqrt{|\Omega_k|}\int_{z_s}^{z_l}\frac{d z'}{E(z')}],\label{2}
\end{eqnarray}
where
\begin{equation}
{\rm sinn}(x)=\begin{cases}
{\rm sin}(x), & \Omega_k<0,  \\
x, & \Omega_k=0 ,\\
{\rm sinh}, & \Omega_k>0,
\end{cases}
\label{3}
\end{equation}
and $E(z)=H(z)/H_0$ is the reduced Hubble parameter at redshift $z$. For convenience, we denote the dimensionless comoving distance $d(0,z_l)$, $d(0,z_s)$, $d(z_l,z_s)$ as $d_l$, $d_s$ and $d_{ls}$, respecitively.

The distance sum rule \citep{rasanen2015new} involves these three dimensionless distances and cosmic curvature $\Omega_k$, which satisfy the following relation
\begin{equation}
\frac{d_{ls}}{d_s}=\sqrt{1+\Omega_k d_l^2}-\frac{d_l}{d_s}\sqrt{1+\Omega_k d_s^2}.  \label{4}
\end{equation}
Obviously, $\Omega_k$ can be derived directly without introducing any cosmological model, once these three dimensionless distances are obtained from observational data. The information of distance ratio $d_{ls}/d_s$ can be derived from the separation of multiple images in SGL observations. Therefore, to obtain the distances $d_l$ and $d_s$ is the first step for the constraint on $\Omega_k$.

In this work, we use the SN Ia (standard candles) observation providing distance information to calibrate the distances $d_l$ and $d_s$ in the SGL systems.

\subsection{Type Ia supernovae}

In this paper, we choose to use the latest SN Ia sample, i.e., the Pantheon sample, from the Pan-STARRS1 Medium Deep Survey. This catalogue consists of 1048 SN Ia data covering the redshift range $0.001<z<2.3$.

Through the relation between comoving dimensionless distance and luminosity distance, we can obtain
\begin{equation}
d(z)=\frac{H_0D_L(z)}{c(1+z)},   \label{5}
\end{equation}
where $D_L$ is the luminosity distance and $c$ is the speed of light.

For an SN Ia, the distance modulus and the luminosity distance $D_L$ are related by $\mu_{\rm th}=5\lg (D_L/{\rm Mpc})+25$ theoretically, and the observed distance modulus is
\begin{eqnarray}
\mu _ { \mathrm { obs } }(z)&=& m_{B}(z)+\alpha \cdot X _ { 1 } - \beta
\cdot \mathcal{C} - M _ { B },
\end{eqnarray}
where $m_{B}$ is the rest frame \textit{B}-band peak magnitude, $X_1$ and $\mathcal{C}$ describe the time stretch of light curve and the supernova color at maximum brightness, respectively. The parameter $M_B$ is the absolute \textit{B}-band magnitude. In general, $\alpha$, $\beta$ and $M_B$ are always as nuisance parameters to be fitted in the distance estimate. To dodge this problem, based on the approach proposed by \citet{Marriner:2011mf}, and by including extensive simulations for correcting the SALT2 light curve fitter, \citet{Kessler:2016uwi} proposed a new method called BEAMS with Bias Corrections (BBC) to calibrate each SN Ia. Therefore, for the Pantheon sample, the stretch-luminosity parameter $\alpha$ and the color-luminosity parameter $\beta$ are calibrated to zero, and then the observed distance modulus is simply expressed as
\begin{equation}
\mu_{\rm obs}(z)=m_B(z)-M_B.
\end{equation}
Once the absolute magnitude of an SN Ia is known, the luminosity distance can be obtained. 

\subsection{Gaussian Process}

\begin{figure}[!htp]
\includegraphics[scale=0.55]{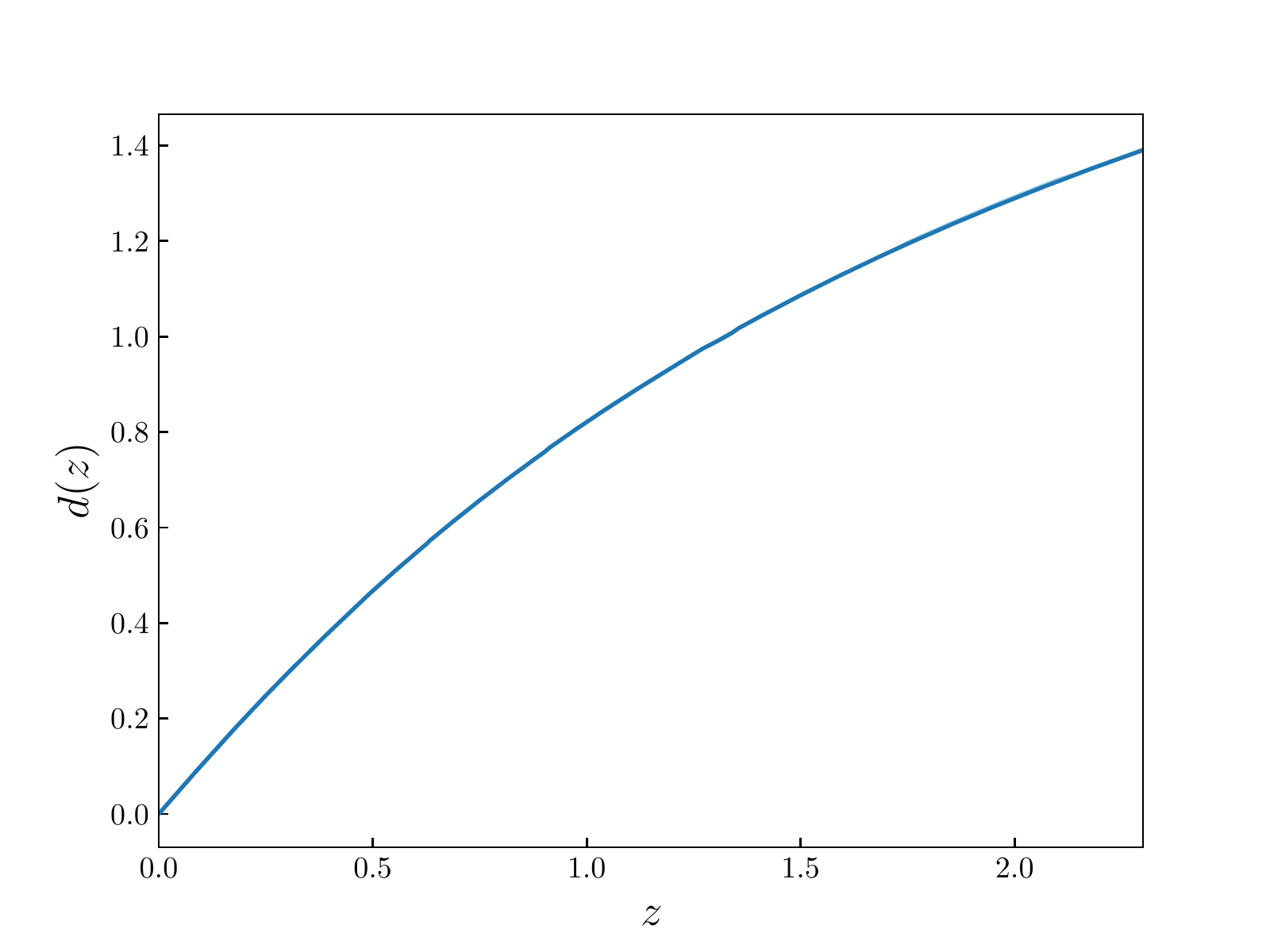}                                                                                                                                               \caption{\label{fig1} Reconstruction of the dimensionless comoving distance $d(z)$ from the SN Ia data. The blue shaded area and blue line represent the $68.3\%$ confidence level errors and best-fit value of the reconstruction by using Gaussian Process regression.}
\end{figure}

The difficulty of calibrating $d_l$ and $d_s$ in SGL system lies in the fact that there is no a one-to-one correspondence between the redshifts of SGL data and SN Ia data. The previous way of treating this problem is to use the polynomial fitting to simulate a smooth distance-redshift curve. {In order to reduce the systematic uncertainty, we use the Gaussian Process (GP) regression \citep{Holsclaw:2010sk,Shafieloo:2012ht,Keeley:2019hmw,Liao:2019qoc,Liao:2020zko} based on \texttt{GPHist} \citep{gphistdoi} Python code first used in \citet{Joudaki:2017zhq}. In our work, we use the GP regression to reconstruct the expansion history $H(z)$, and we choose the standard squared exponential covariance function to connect the values between different points $s$:}
\begin{eqnarray}
\langle\gamma(s_1)\gamma(s_2)\rangle&=&\sigma_f^2e^{-(s_1-s_2)^2/(2\ell^2)},\label{8}
\end{eqnarray}
{where $\sigma_f$ and $\ell$ are hyperparameters, characterizing the amplitude of deviation and the correlation scale, respectively. The hyperparameters are important for both physical insight and error control, and they can be determined by fitting data. For a more accurate reconstruction of $H(z)$, we control the dynamical range by defining the variable $\gamma(s)=\ln[H(z)_{\rm fid}/H(z)]$ and $s(z)=\ln(1+z)/\ln(1+z_{\rm max})$ as shown in \citet{Keeley:2019hmw}, where $z_{\rm max}=2.3$ is the highest redshift in the Pantheon dataset, and $H(z)_{\rm fid}$ is from the best-fit $\Lambda$CDM model. With $H(z)$ in hand, we can calculate angular diameter distances (or luminosity distances) by using Eq.~(\ref{2}).

{The Pantheon compilation \citep{scolnic2018complete} provides distance moduli $\mu_{\rm obs}$ up to an absolute magnitude $M_B$ for 1048 SN Ia, described by a vector $\hat{{\bm X}}$, where $\hat{X}^i=\mu_{\rm obs}^i+M_B$ with the hat denoting the Pantheon data, and its covariance matrix ${\bm C}$ (including both statistical and systematic uncertainties). Therefore, we can infer the luminosity distances $D_L$ through establishing the likelihood $\mathcal{L}$ defined by $-2 \ln \mathcal{L} = (\hat{{\bm X}}- {\bm X})^T {\bm C}^{-1}(\hat{{\bm X}} - {\bm X})$. The model vector ${\bm X}$ is composed of $\mu_{\rm th}^i=5\lg (D_L(z_i)/{\rm Mpc})+25$, where $z_i$ is the $i$-th SN Ia redshift and $D_L$ is generated by the GP regression. The value of the absolute magnitude $M_B$ can be estimated by }
\begin{eqnarray}
M_B=\sum_{i=1}^{1048} \frac{\hat{X}^i-\mu_{\rm th}^i}{(\Delta \hat{X}^i)^2}/\sum_{i=1}^{1048}\frac{1}{(\Delta \hat{X}^i)^2},
\end{eqnarray}
where $\Delta \hat{X}$ is the error of $\hat{X}$. Finally, we can use Eq.~(\ref{5}) to reconstruct the dimensionless comoving distances. Following the motivation of this work, to constrain $\Omega_k$ by using low-redshift observations with a cosmological model-independent method, in Eq.~(\ref{5}) we consider the latest local measurement of the Hubble constant, with the central value of $H_0=74.03 \rm{~km~s}^{-1} \rm{~Mpc}^{-1}$, from the Cepheid-supernova distance ladder \citep{riess2019large}. The reconstructed dimensionless comoving distance versus redshift is shown in Fig.~\ref{fig1}.}


\subsection{Strong gravitational lensing systems}

With more and more strong gravitational lensing systems observed, the SGL observation has become an important tool to explore extragalactic astronomy \citep{ofek2003redshift,cao2016limits} and cosmology \citep{biesiada2010cosmic,biesiada2011dark,cao2011testing,cao2012constraints,cao2014cosmic,li2016comparison}. Recently, \citet{Chen:2018jcf} compiled the largest sample of SGL to date from the following surveys: Sloan Lens ACS survey \citep{bolton2006sloan}, BOSS Emission-Line Lens Survey \citep{brownstein2011boss}, Lenses Structure and Dynamics survey \citep{treu2004massive}, CFHT Strong Lensing Legacy Survey \citep{cabanac2007cfhtls}. This newly compiled SGL sample includes 161 galaxy-scale strong lensing systems, for which the redshift range of lenses is $0.0624\leq z_l \leq 1.004$ and the redshift range of sources is $0.197\leq z_s \leq 3.595$. Due to the fact that the maximum redshift of SN Ia used to calibrate $d_s$ is about $2.3$, the SGL systems with $z_s>2.3$ should be excluded so that there are 135 SGL systems in our study actually.

In order to explore whether there is a bias for different lenses masses, we divided the SGL sample into three sub-samples based on the center velocity dispersion of intervening galaxies, and they are low-mass galaxies ($\sigma_{\rm{ap}}\leq 200~ {\rm km/s}$), high-mass galaxies ($\sigma_{\rm{ap}}> 300~{\rm km/s}$), and intermediate-mass galaxies ($200~{\rm km/s}<\sigma_{\rm{ap}}\leq300 ~{\rm km/s}$). From the spectroscopic data, the velocity dispersion $\sigma_{\rm{ap}}$ is measured within a circular aperture with the angular radius $\theta_{\rm{ap}}$, which should be transformed to velocity dispersion within a circular aperture of radius $R_{\rm{eff}}/2$ (half of the effective radius) according to the aperture correction formula \citep{Jorgensen:1995zz},
\begin{equation}
\sigma_0=\sigma_{\rm{ap}}(\frac{\theta_{\rm{eff}}}{2\theta_{\rm{ap}}})^{\eta}, \label{9}
\end{equation}
and here we adopt the value $\eta=-0.066$ \citep{Cappellari:2005ux,Chen:2018jcf}. The error of $\sigma_0$ could be obtained by using the error transfer formula from the uncertainty of $\sigma_{\rm{ap}}$. In addition, we will take the fractional uncertainty of the Einstein radius at the level of $5\%$ in the following analysis.

For strong gravitational lensing systems, although the properties of galaxies as lenses are not fully understood in detail, two models are commonly used to describe the mass distribution of galaxies, i.e., the singular isothermal sphere (SIS) model and the singular isothermal ellipsoid (SIE) models \citep{koopmans2006sloan}. In this paper, to investigate the effects of different lens models on the constraints on $\Omega_k$, three lens models widely used are considered.

(i) SIS model: Assuming that the mass distribution of lens galaxies is described approximately by the SIS model, the distance ratio can be expressed as:
\begin{equation}
\frac{d_{ls}}{d_s}=\frac{c^2\theta_E}{4\pi\sigma_{\rm{SIS}}^2},  \label{10}
\end{equation}
where $\sigma_{\rm{SIS}}$ is the velocity dispersion of lens galaxy reflecting the total mass of the lens. In order to consider the uncertainties between the observed stellar velocity dispersion and the SIS model, as well as other systematic effects, a phenomenological coefficient $f_E$ is introduced as $\sigma_{\rm{SIS}}=f_E\sigma_{0}$, where $\sigma_{0}$ is the observed velocity dispersion. Here, we take $f_E$ as a free parameter with $0.8<f_E^2<1.2$ \citep{ofek2003redshift}.

(ii) Power-law spherical model: Considering a more general lens model, we assume that the total mass density profile of the lensing galaxy is characterized by the spherically symmetric power-law distribution $\rho\sim r^{-\gamma}$, where $r$ is the spherical radius from the center of the lensing galaxy. According to the spherical Jeans equation, the distance ratio can be expressed as \citep{koopmans2006sloan}:
\begin{equation}
\frac{d_{ls}}{d_s}=\frac{c^2\theta_E}{4\pi \sigma_{\rm{ap}}^2}\(\frac{\theta_{ap}}{\theta_E}\)^{2-\gamma} f^{-1}(\gamma), \label{11}
\end{equation}
where
\begin{equation}
f(\gamma)=-\frac{1}{\sqrt{\pi}}\frac{(5-2\gamma)(1-\gamma)}{3-\gamma}\frac{\Gamma(\gamma-1)}{\Gamma(\gamma-3/2)}\[\frac{\Gamma(\gamma/2-1/2)}{\Gamma(\gamma/2)}\]^2.  \label{12}
\end{equation}
As one can see, the power-law model will be reduced to the SIS model when $\gamma=2$. Some studies suggested that the early-type galaxies may be evolving with redshift \citep{Bolton:2012uh,Sonnenfeld:2013xga,cao2016limits}. Therefore, we consider the power-law index is varying with redshift $\gamma(z)=\gamma_0+\gamma_1 z_l$ to account for the possible evolution of mass density profile, where $\gamma_0$ and $\gamma_1$ are two free parameters in our analysis.

(iii) Extended power-law model: Considering the influence of dark matter on the mass distribution, we assume the luminosity density profile $\nu(r)$ is different from the total-mass density profile $\rho(r)$. Their expressions are
\begin{equation}
\rho(r)=\rho_0 {\(\frac{r}{r_0}\)}^{-\alpha},~~~ \nu(r)=\nu_0 {\(\frac{r}{r_0}\)}^{-\delta}, \label{13}
\end{equation}
where $r$ is the spherical radial coordinate from the lens center, and $\alpha$ and $\delta$ are two free parameters. Based on the above two equations, the observational distance ratio can be expressed as:
\begin{align}
\frac{d_{ls}}{d_s}=&\frac{c^2\theta_E}{2\sigma_0^2\sqrt{\pi}}\frac{3-\delta}{(\xi-2\beta)(3-\xi)}\(\frac{\theta_{\rm{eff}}}{\theta_E}\)^{2-\alpha}   \n\\
\times&\[\frac{\lambda(\xi)-\beta\lambda(\xi+2)}{\lambda(\alpha)\lambda(\delta)}\],\label{14} \end{align}
where $\xi=\alpha+\delta-2$ and $\lambda(x)=\Gamma(\frac{x-1}{2})/\Gamma(\frac{x}{2})$. The parameter $\beta$ characterizes the anisotropic distribution of three-dimensional velocity dispersion. Based on the well-studied sample of nearby elliptical galaxies \citep{Gerhard:2000ck}, we treat $\beta$ as a nuisance parameter and marginalize over it with a Gaussian prior $\beta=0.18\pm0.13$. In the case of $\alpha=\delta=2$ and $\beta=0$, the extended power-law model is reduced to the SIS model.

We constrain $\Omega_k$ by using the \texttt{emcee} Python module \citep{foreman2013emcee} based on the Markov-chain Monte Carlo (MCMC) method to maximize the likelihood $\mathcal{L}\varpropto e^{-\chi^2 /2}$. The $\chi^2$ function is defined as
\begin{equation}
\chi^2(\textbf{p},\Omega_k)=\sum\limits^{135}_{i=1}\frac{(D_{\rm{th}}(z_i;\Omega_k)-D_{\rm{obs}}(z_i;\rm{\textbf{p}}))^2}{\sigma_{D}(z_i)^2}, \label{15}
\end{equation}
where $D=d_{ls}/d_s$, and ${\bf p}$ represents the parameters of different SGL system models. Here, $\sigma_{D}$ is the uncertainty of $D$, which is contributed by both uncertainties from the observation of SGL and from the distance calibration by the SN Ia data. {To be specific, as shown in Fig.~\ref{fig1}, the dimensionless comoving distance with 1$\sigma$ uncertainty has been reconstructed from the SN Ia data. The central values and errors of $d_l$ and $d_s$ used in Eq.~(\ref{4}) can be obtained from the median and size of the band of GP reconstruction at the same redshift, respectively. Then, the value of $\sigma_{\rm{SN}}^2$ could be obtained by applying the error transfer formula to Eq.~(\ref{4}).} We assume that they are uncorrelated and thus we have $\sigma_D^2=\sigma_{\rm{SGL}}^2+\sigma_{\rm{SN}}^2$. Additionally, in order to realize Eq. (\ref{4}), the conditions of $1+\Omega_kd_s^2\geq0$ and $1+\Omega_kd_l^2\geq0$ must be held. According to the maximum distance of SGL sample we have used, we take a prior range for the cosmic curvature in MCMC as $\Omega_k\geq-0.5$.

\section{Results and discussion}\label{sec3}

\subsection{SIS model}
\begin{figure*}[!htp]
\includegraphics[scale=0.38]{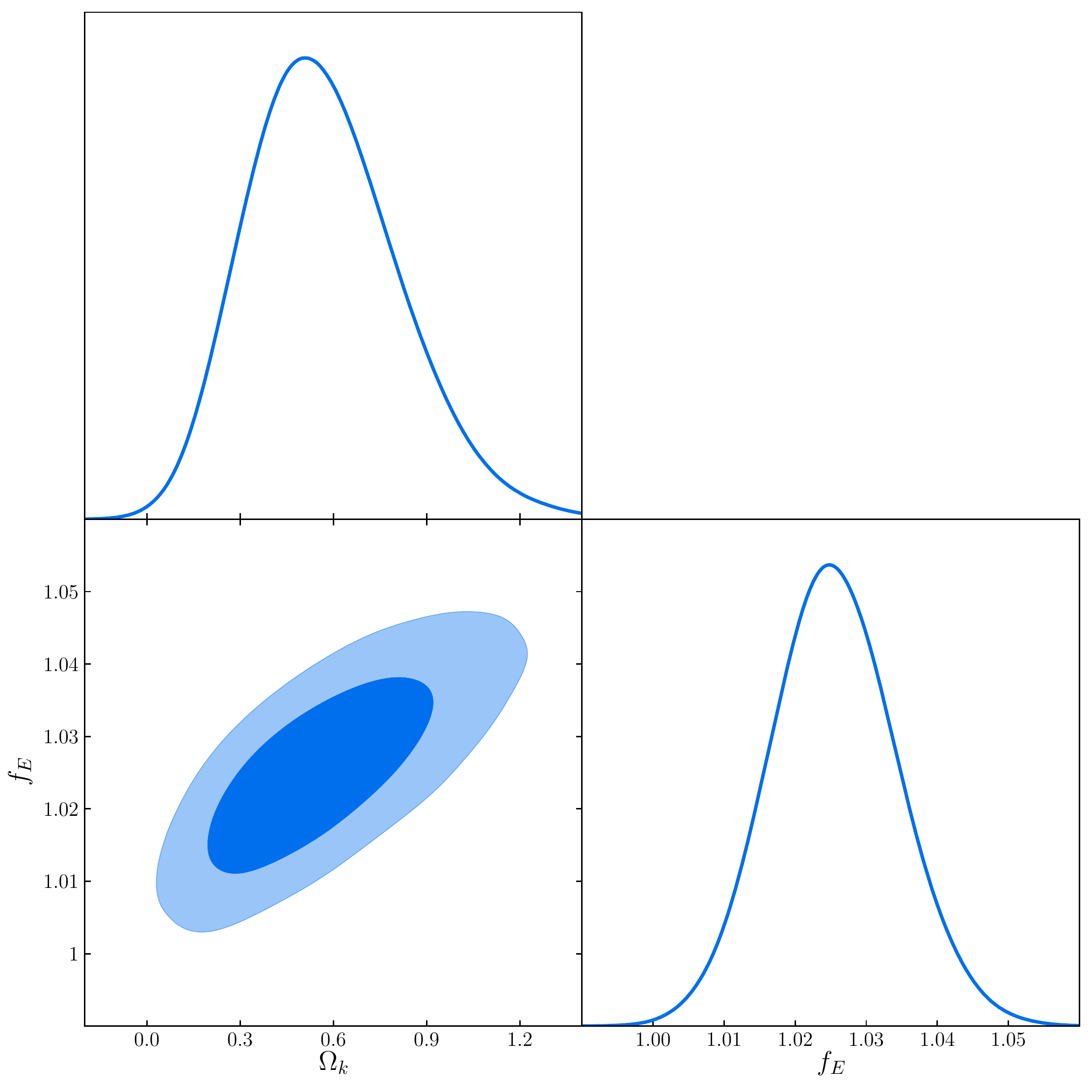}
\includegraphics[scale=0.38]{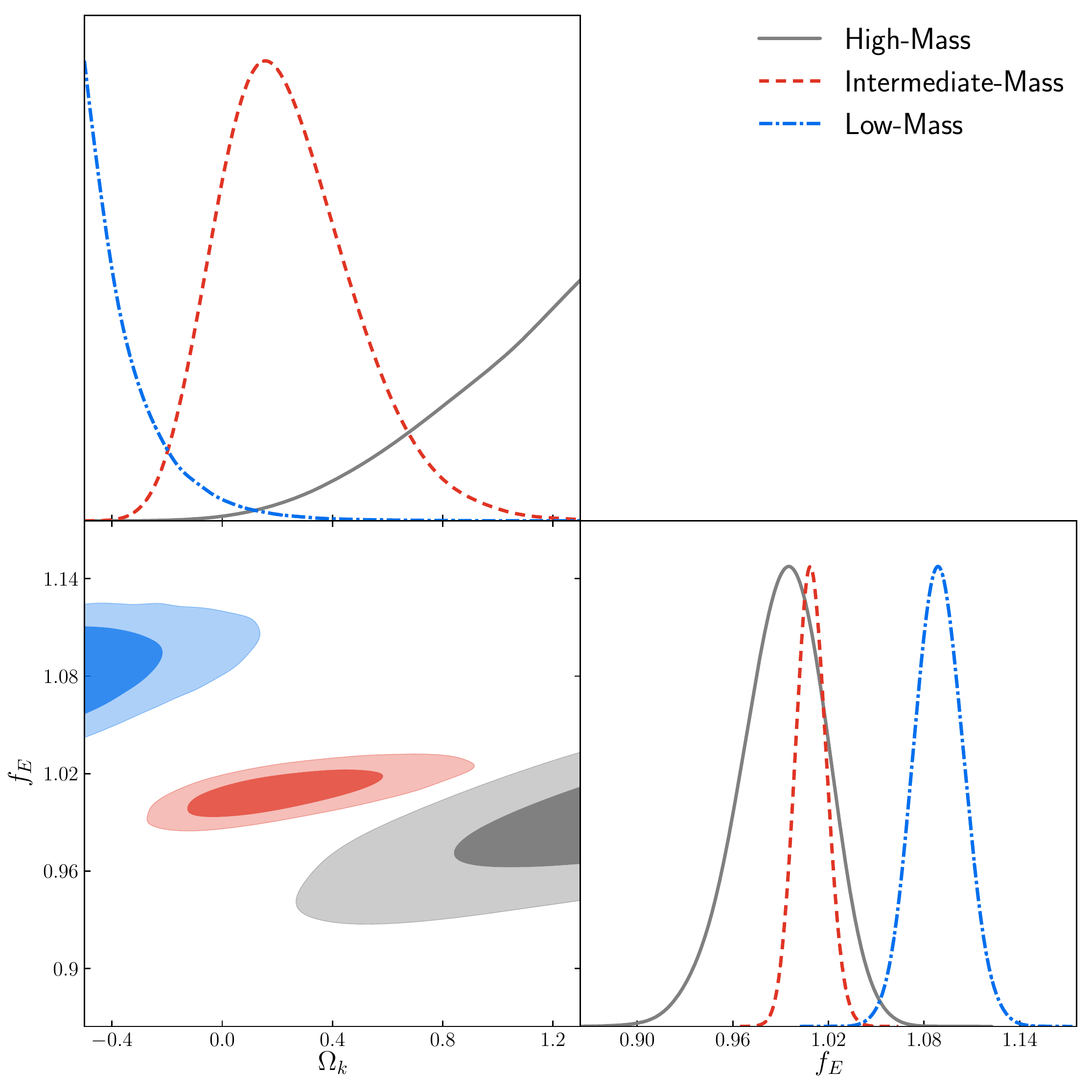}
\centering
\caption{\label{fig2}Left: The 1D and 2D posterior distributions for the cosmic curvature $\Omega_k$ and the SGL system parameter ($f_E$) from the full sample in the SIS model. Right: Constraints from the three subsamples of the SGL full sample classified according to their center velocity dispersions.}
\end{figure*}

The constraints on $\Omega_k$ and $f_E$ of the SIS model are shown in Fig.~\ref{fig2} and Table~\ref{tab1}. {According to the constraint results from the full data, a spatially flat universe is not preferred at the $2\sigma$ level.} Although the constraint uncertainty of $\Omega_k$ is not as small as that from the Planck observation, it is still rather meaningful because our results are obtained by using a cosmological model-independent method and the low-redshift observations.

To investigate the impact of the lens mass on the results, we show the constraints on $\Omega_k$ and $f_E$ for the the three sub-samples based on the center velocity dispersion of intervening galaxies in the right panel of Fig. \ref{fig2}. We find that the constraints on $\Omega_k$ from three subsamples are different from each other. This tension further confirms the conclusion of \cite{qi2018distance} and \cite{li2018curvature} that it needs to treat separately subsamples with different velocity dispersions. Moreover, it is interesting to see that the constraints on $\Omega_k$ obtained by low-mass, intermediate-mass, and high-mass subsamples correspond coincidently to closed, flat, and open universe cases, respectively. Therefore, if the Planck result of a zero value of $\Omega_k$ is true (for which the $\Lambda$CDM model is assumed), then we find that only the intermediate-mass galaxies are suitable to be described by the SIS model. Actually, we also find that the uncertainty of $\Omega_k$ from the intermediate-mass subsample is almost as the same as that obtained from the full data.

For the parameter $f_E$, we find that the constraints from the three subsamples are also different, and we can see that the results of the high-mass subsample and the intermediate-mass subsample are in good agreement with the SIS model ($f_E=1$), while the result of the low-mass subsample actually excludes the SIS model at the 2$\sigma$ CL. Thus, the tensions for $\Omega_k$ and $f_E$ from different subsamples imply that it is not reasonable to use the uniform SIS model to characterize all the lenses.



\begin{deluxetable}{ccccc}[hbp]
\tabletypesize{\scriptsize}
\tablecaption{The best-fit values of the cosmic curvature $\Omega_k$ and the SGL system parameter ($f_E$) at the $68\%$ confidence level in the SIS model for all the cases.\label{tab1}}
\tablewidth{0pt}
\tablehead{
& \colhead{low-mass} & \colhead{intermediate-mass} & \colhead{high-mass} & \colhead{full sample}
}
\startdata
\hline
$\Omega_k$& $<-0.316$ & $0.24^{+0.19}_{-0.28}$ & $>1.26$ & $0.57^{+0.20}_{-0.28}$ \\
\hline
$f_E$ & $1.089\pm0.016$ & $1.0091\pm0.0097$ & $0.991_{-0.024}^{+0.029}$ & $1.0255\pm0.0090$ \\
\hline
\enddata
\end{deluxetable}

\subsection{Power-law lens model}

\begin{figure*}[!htp]
\includegraphics[scale=0.38]{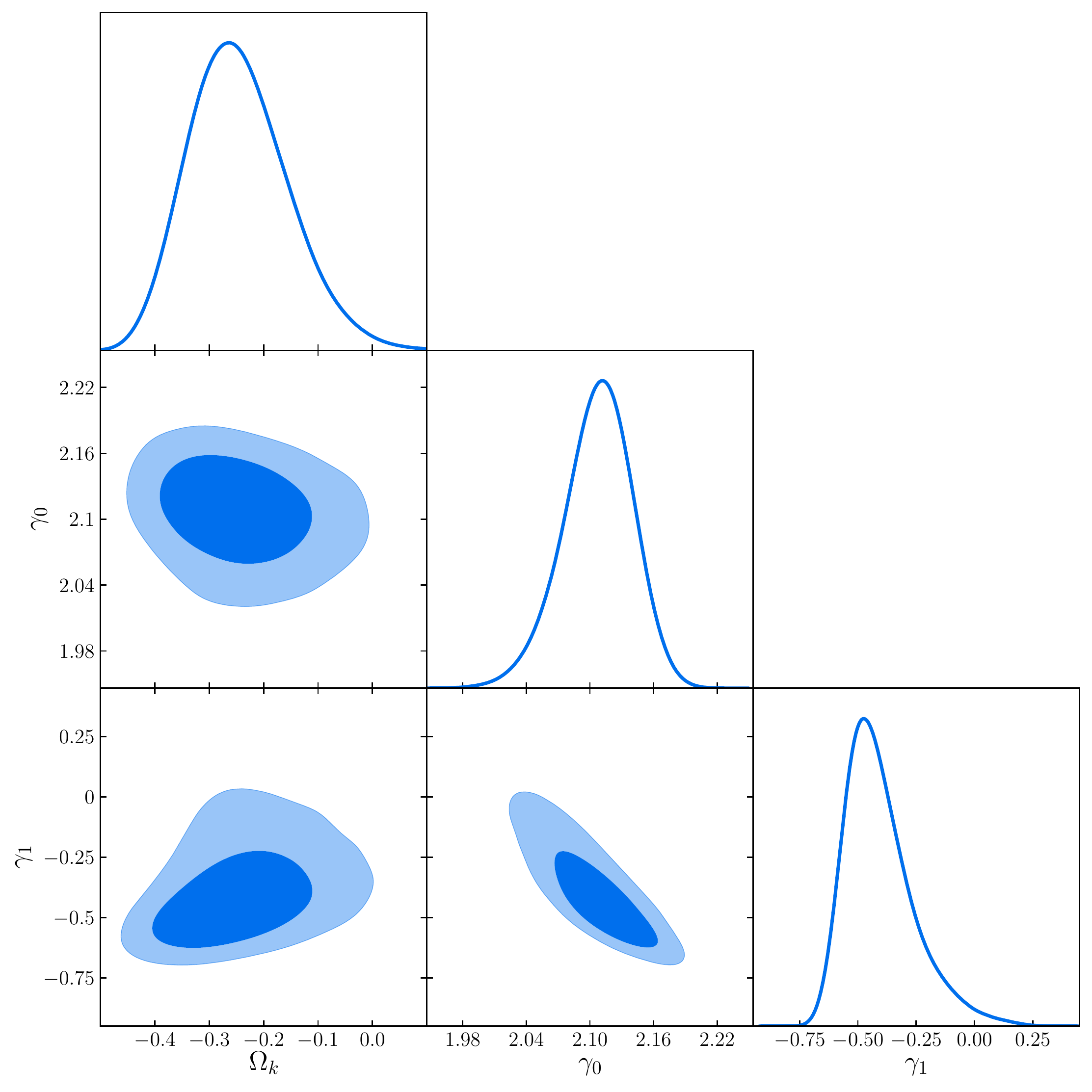}
\includegraphics[scale=0.38]{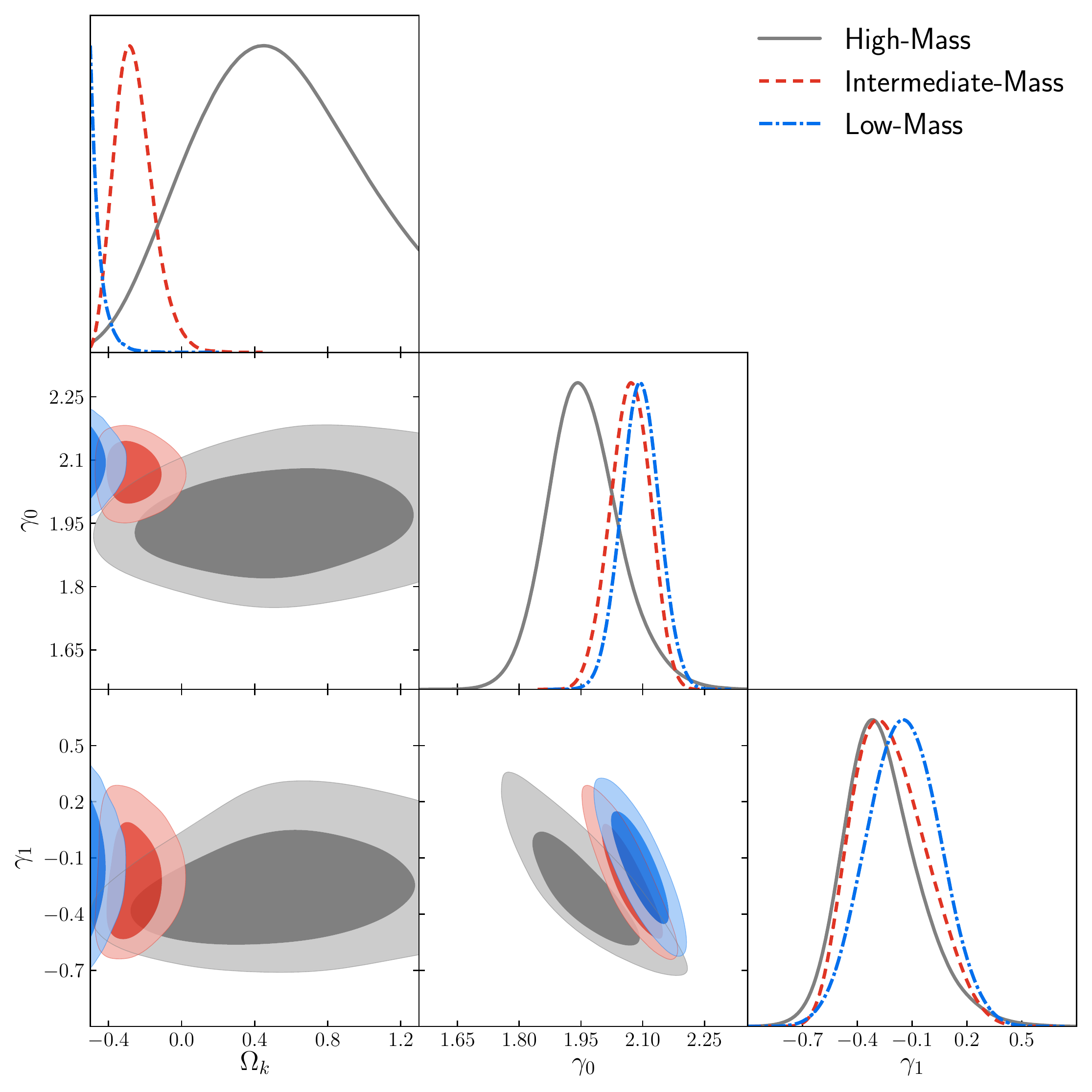}
\centering
 \caption{\label{fig3}Left: The 1D and 2D posterior distributions for the cosmic curvature $\Omega_k$ and the SGL system parameters ($\gamma_0,\gamma_1$) from the full sample in the power-law lens model. Right: Constraints from the three subsamples of the SGL full sample classified according to their center velocity dispersions.}
\end{figure*}

For the spherically symmetric power-law lens model, we show the constraint results in Fig.~\ref{fig3} and Table~\ref{tab2}. The constraint result of $\Omega_k$ from the full sample is $\Omega_k=-0.246^{+0.078}_{-0.100}$ at the $1\sigma$ CL, indicating that a flat universe is excluded by this analysis. Compared with the results of the SIS model, we find that the choice of the lens models will lead to rather different results for constraints on $\Omega_k$.

For the impact of the subsamples, we find that in this case the intermediate-mass subsample excludes the flat universe at the more than 2$\sigma$ level, rather than strongly supporting the flat universe as in the SIS model. Actually, in this case, the high-mass subsample supports the flat universe.

We note that the power-low lens model will be reduced to the SIS model when $\gamma_0=2$ and $\gamma_1=0$. By using the full sample, we obtain the results $\gamma_0=2.107^{+0.036}_{-0.029}$ and $\gamma_1=-0.405^{+0.090}_{-0.170}$. Thus, we can see that the full-sample results do not support the power-low lens model being reduced to the SIS model at the 1$\sigma$ level, especially the results of $\gamma_1$ has rule out zero at 2$\sigma$ confidence level. This suggests that the total density profile of early-type galaxies may be evolving with cosmic time slightly.

For the case of subsamples, we find that the tensions for the parameters determined from different subsamples are also rather evident. But we can see that in this lens model the results of intermediate-mass subsample and low-mass subsample are consistent in some extent. From the fit results of $\gamma_0$ and $\gamma_1$ (see Table~\ref{tab2}), we find that none of the three subsamples supports the power-low model being reduced to the SIS model at the 1$\sigma$ level.



\begin{deluxetable}{ccccc}[hbp]
\tabletypesize{\scriptsize}
\tablecaption{\label{tab2}The best-fit values of the cosmic curvature $\Omega_k$ and the SGL system parameters ($\gamma_0,\gamma_1$)  at the $68\%$ confidence level in the power-law lens model for all the cases.}
\tablewidth{0pt}
\tablehead{
& \colhead{low-mass} & \colhead{intermediate-mass} & \colhead{high-mass} & \colhead{full sample}
}
\startdata
\hline
$\Omega_k$&$<-0.445$& $-0.259^{+0.082}_{-0.12}$ & $0.59^{+0.39}_{-0.59}$ & $-0.246^{+0.078}_{-0.100}$ \\
\hline
$\gamma_0$& $2.094\pm0.046$ & $2.069^{+0.051}_{-0.046}$ & $1.958_{-0.094}^{+0.076}$ & $2.107^{+0.036}_{-0.029}$ \\
\hline
$\gamma_1$& $-0.15\pm0.20$ & $-0.22^{+0.17}_{-0.24}$ & $-0.26_{-0.23}^{+0.16}$ & $-0.405^{+0.090}_{-0.170}$ \\
\hline
\enddata
\end{deluxetable}

\subsection{Extended power-law lens model}

\begin{figure*}[!htp]
\includegraphics[scale=0.38]{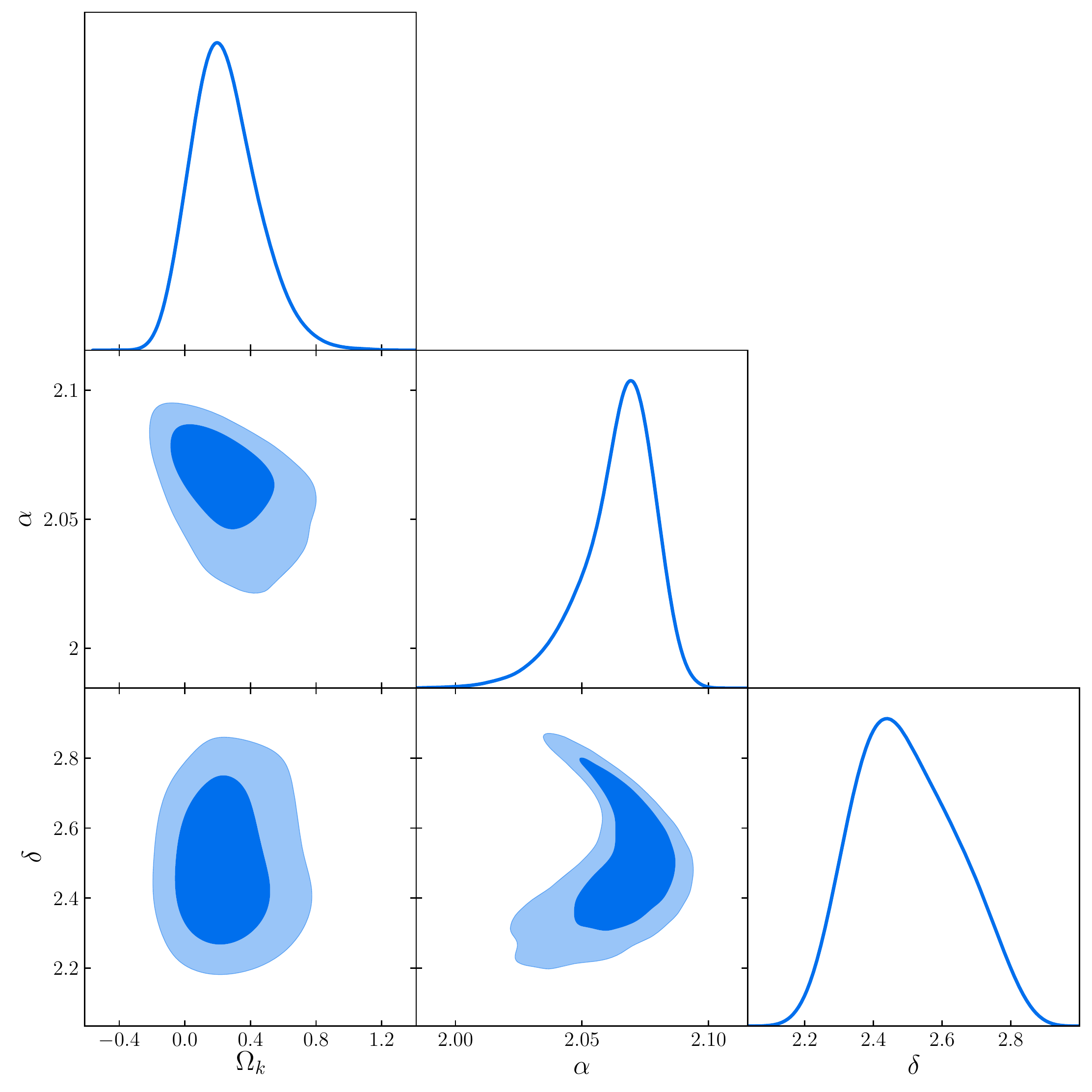}
\includegraphics[scale=0.38]{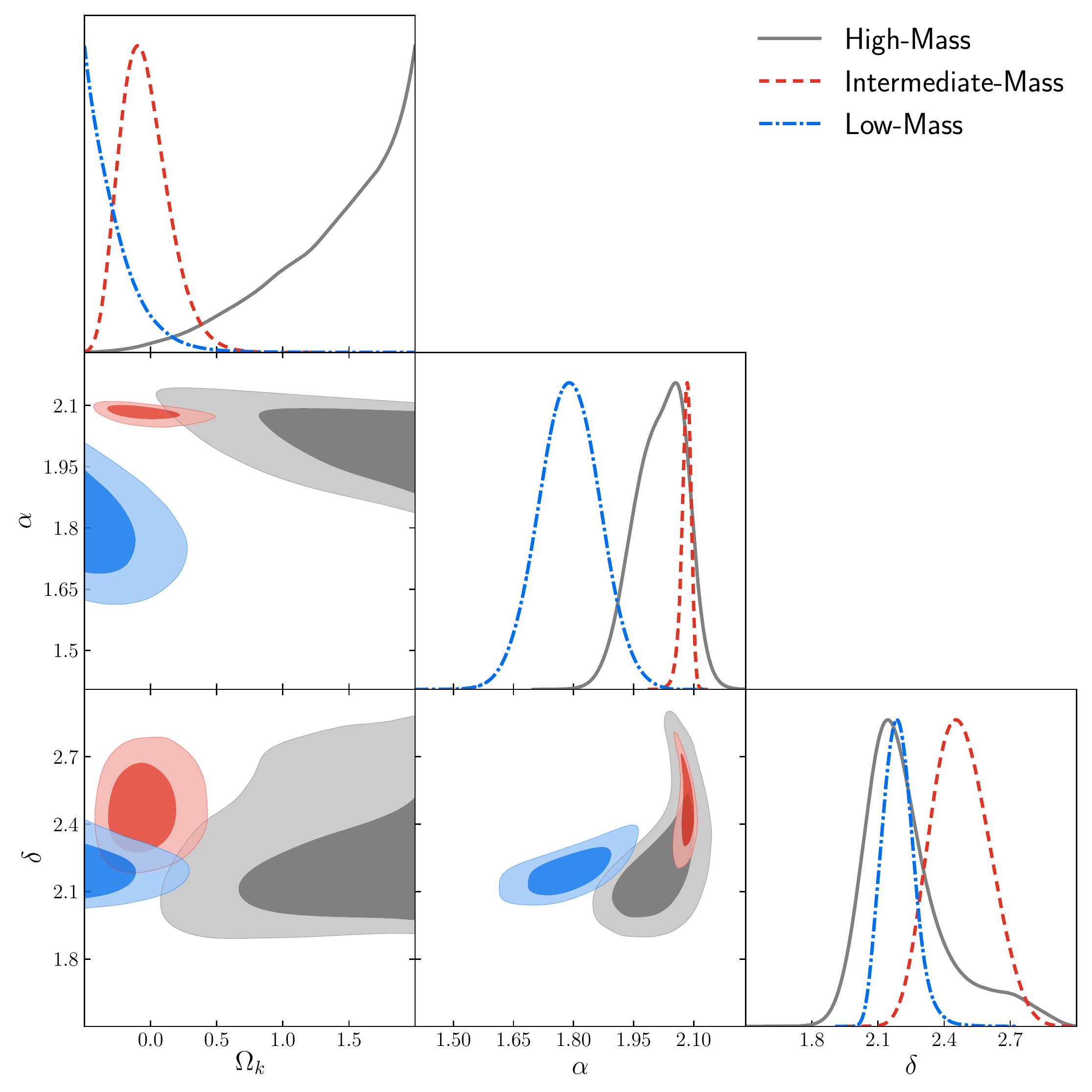}
\centering
 \caption{\label{fig4} Left: The 1D and 2D posterior distributions for the cosmic curvature $\Omega_k$ and the SGL system parameters ($\alpha,\delta$) from the full sample in the power-law lens model. Right: Constraints from the three subsamples of the SGL full sample classified according to their center velocity dispersions.}
\end{figure*}

For the extended power-law lens model, we show the constraint results in Fig.~\ref{fig4} and Table~\ref{tab3}. We find that a flat universe is favored by using the full sample when this model is considered (see the left panel of Fig.~\ref{fig4}). The difference between  the luminosity mass density profile and the total mass density profile ($\alpha\neq\delta\neq 2$) reveals that the effect of dark matter in the early-type galaxies is not negligible, and its mass density distribution does not obey the distribution law of baryons. Therefore, the issue of mass density profile in the lens galaxies actually needs to be further investigated.

The constraint results from the subsamples are shown in the right panel of Fig.~\ref{fig4}. The great tensions for the constraints on the parameters from different subsamples can directly be seen from this figure. We find that, similar to the case of the SIS model, the low-mass, intermediate-mass, and high-mass subsamples prefer a closed, flat, and open universes, respectively, in this case. We find that, for the parameter $\alpha$, the results from the high-mass and interediate-mass subsamples are consistent in some extent, and for the parameter $\delta$, the results from the high-mass and low-mass subsamples are consistent in some extent. However, on the whole, the constraint results from the three subsamples are in great tensions.



\begin{deluxetable}{ccccc}[hbp]
\tabletypesize{\scriptsize}
\tablecaption{\label{tab3} The best-fit values of the cosmic curvature $\Omega_k$ and the SGL system parameters ($\alpha,\delta$) at the $68\%$ confidence level in the extended power-law lens model for all the cases.}
\tablewidth{0pt}
\tablehead{
& \colhead{low-mass} & \colhead{intermediate-mass} & \colhead{high-mass} & \colhead{full sample}
}
\startdata
\hline
$\Omega_k$& $<-0.244$ & $-0.04^{+0.14}_{-0.21}$ & $>1.24$ & $0.25^{+0.16}_{-0.23}$ \\
\hline
$\alpha$& $1.790\pm0.075$ & $2.081^{+0.014}_{-0.009}$ & $2.016^{+0.074}_{-0.053}$ & $2.064^{+0.018}_{-0.009}$ \\
\hline
$\delta$& $2.193_{-0.077}^{+0.061}$ & $2.47^{+0.12}_{-0.14}$ & $2.231^{+0.092}_{-0.220}$ & $2.50_{-0.18}^{+0.13}$ \\
\hline
\enddata
\end{deluxetable}

\section{Conclusion}\label{sec4}

In this paper, with a newly compiled SGL sample including 161 data and the latest Pantheon SN Ia sample including 1048 data, we obtained cosmological model-independent constraints on the cosmic curvature parameter by applying the distance sum rule. To reduce the systematic uncertainty, we use the Gaussian Process regression to reconstruct a smooth distance-redshift curve from the SN Ia data straightforwardly without any parametric assumption and then use the result to calibrate $d_l$ and $d_s$ in the SGL systems. For the lens models, we considered the SIS model, the power-law mass profile lens model, and the extended power-law lens model taking into account the different profiles in total and luminous masses of the lens. Furthermore, we also investigated the constraints on $\Omega_k$ in three subsamples based on the lens velocity dispersion: low-mass galaxies with $\sigma_{\rm{ap}}\leq 200$ km/s, intermediate-mass galaxies with $200~ \rm{km/s}<\sigma_{\rm{ap}}\leq 300$ km/s, and high-mass galaxies with $\sigma_{\rm{ap}}>300$ km/s.

Compared with the results obtained in the previous work \citep{rasanen2015new,xia2017revisiting,li2018curvature,qi2018distance,wang2020machine} using the same method as our analysis, we got some different results. For instance, in \cite{rasanen2015new} and \cite{xia2017revisiting}, a prior of $\Omega_k\geq -0.1$ from the CMB observation is considered in their analysis, which prevents the constraint values of $\Omega_k$ from being more negative, but even though such a prior is considered, the constraints are still weak and a flat universe is only slightly favored.
As expected, when this prior is removed, \citet{li2018curvature} found that a spatially closed universe is favored, and they suspected that there are some unknown systematics leading to a bias in the estimation of the cosmic curvature parameter $\Omega_k$ with SGL observations. In this paper, without the CMB prior, we reinvestigated this issue with larger samples of SN Ia and SGL observations considering more general models for the lens density profile and more complex data classifications. We found that in the power-law lens model a closed universe is preferred, and in the SIS model a open universe is preferred. Only the extended power-law lens model a flat universe can be supported well. Therefore, the larger sample of SGL systems could reliably improve the constraints on $\Omega_k$.

We also found that there are two factors significantly influencing the constraints on the cosmic curvature parameter, i.e., the choice of lens models and the classifications of SGL data. The uncertainty of $\Omega_k$ obtained from the latest observations is about 0.2, which is large compared to the result given by the Planck observation, but we emphasize that such a result is obtained by using a cosmological model-independent method and only low-redshift observations. Considering the lens models, we find that the SIS model ($f_E=1$) is well consistent with results from intermediate-mass and high-mass galaxies. In the framework of power-law mass model, the values of lens model parameters ($\gamma_0, ~\gamma_1$) obtained from high-mass subsample is barely consistent with those in the SIS model ($\gamma_0=2, ~\gamma_1=0$), but from others are not which implies that the total density profile of early-type galaxies may be evolving with cosmic time slightly. When considering the luminosity density profile to be different from the total mass density profile, in the extended power-law lens model, the constraints on the lens model parameters ($\alpha,~\delta$) from different subsamples are inconsistent with each other, revealing the possible difference between mass density distributions of dark matter and luminous baryons in galaxies with different masses.

\section*{Acknowledgments}
We would like to thank Shuo Cao and Yun Chen for helpful discussions. This work is supported by the National Natural Science Foundation of China (Grants No. 11975072, No. 11875102, No. 11835009, and No. 11690021), the Liaoning Revitalization Talents Program (Grant No. XLYC1905011), the National Program for Support of Top-Notch Young Professionals, and the Fundamental Research Funds for the Central Universities of China (Grants No. N2005030, No. N180503014, and No. N182410008-1).

\bibliography{ref2}
\bibliographystyle{aasjournal}

\end{document}